\scrollmode
\documentclass[english,12pt]{article}
          \usepackage{babel}
          \usepackage[T1]{fontenc}
          \usepackage[latin2]{inputenc}
          \usepackage{pslatex}
\input epsf

\begin{document}

\voffset -0.5in
\title{An evolutionary considerations for V228 from 47 Tuc}

\author{Marek J. Sarna}
\date{ N. Copernicus Astronomical Center,
       Polish Academy of Sciences,\\
       Bartycka Str. 18, 00-716 Warsaw, Poland.\\
       e-mail: sarna$@$camk.edu.pl}

\maketitle

\begin{abstract} 
We perform evolutionary calculations of binary stars to find
progenitors of system with parameters similar to the eclipsing
binary system V228. We show that a V228 binary system may be formed
starting with an initial binary system which has a low main sequence
star as an accretor. The initial parameters for the evolutionary
model are as follow: $M_{1,i} = 0.88~M_\odot $, $M_{2,i} =
0.85~M_\odot $, $P_i=1.35 ~$days, $f_1$=0.05, $f_2$=4.65 and Z=0.006
([Fe/H]=--0.67). We also show that the best fitting model implies
loss of about 50 per cent of initial total orbital momentum but only
5 per cent of initial total mass. The less massive component have a
small helium core of mass 0.12--0.17$~M_\odot $ and exchange mass in
the nuclear time scale.
\end{abstract}


\section{Introduction}   

Blue straggler (BS) stars are defined by their location on the
color--magnitude diagram. These star lie above the main--sequence
turnoff region, a region where, if the BS's had been normally single
stars, they should already have evolved away from the main sequence.
The eclipsing binary V228 was discovered by Ka\l u\.zny et al.
(1998) during a survey for variable stars in the field of the
globular cluster 47 Tuc, and classify as the eclipsing BS. Authors
found an orbital period of $\rm P = 1.1504$ days. On the
colour--magnitude diagram of the cluster, the variable occupies a
position near the top of BS sequence. Ka\l u\.zny et al. (2007)
concluded that, V228 has the same proper motion and radial velocity
as 47 Tuc and is located at the cluster distance -- member of the
globular cluster 47 Tuc. However, an analysis of Ka\l u\.zny et al.
(2007) shows that V228 is a semi--detached Algol--type binary with
the less massive component filling its Roche lobe. We know several
system of such type: RY Aqr, S Cnc, R CMa, AS Eri (see Table 2).

In this paper we report results of an evolutionary considerations
and their comparision with absolute parameters of V228 determined
from observations (Ka\l u\.zny et al 2007).

\section{What do we know about V228 and 47 Tuc?}

The absolute parameters of V228 from Ka\l u\.zny et al. (2007)
spectroscopic and photometric analysis are given in Table 1.

\begin{table*}
\begin{center}
\begin{tabular}{lccc}
\multicolumn{4}{l}{Table 1 ~~Parameters of primary and secondary of V228} \\
\hline
\multicolumn{1}{l}{} & mass & radius & luminosity \\
\multicolumn{1}{l}{} & [$\rm M_\odot$] & [$\rm R_\odot$] & [$\rm L_\odot$] \\
\hline \multicolumn{4}{l}{}\\
{\bf Primary:} & $\rm M_1: 1.512\pm0.022$ & $\rm R_1:
1.357\pm0.019$ & $\rm L_1: 7.02\pm 0.50 $ \\
{\bf Secondary:} & $\rm M_2: 0.20\pm0.007$ & $\rm R_2: 1.238\pm
0.013$ & $\rm L_2: 1.57\pm 0.09 $ \\
\multicolumn{4}{l}{}\\ \hline \multicolumn{4}{l}{}\\

\multicolumn{1}{l}{Orbital period} & $\rm P = 1.150686$ & & \\
\multicolumn{4}{l}{}\\\hline
\end{tabular}
\end{center}
\end{table*}

For globular cluster 47 Tuc we know that age of the cluster is
between 10 and 14 Gyr (Gratton et al. 2003, VandenBerg et al. 2006,
Ka\l u\.zny et al. 2007). Using VandenBerg et al. (2006) isochrones
for the age of 14 Gyr implies turnoff masses of 0.868 and 0.852$\rm
M_\odot $ for [Fe/H]= -0.606 and -0.707, respectively. We assuming
metalicity value from Alves--Brito et al. (2005) [Fe/H]=--0.67
(Z=0.006).

\section{Evolutionary model}

While calculating evolutionary models of binary stars, we must
take into account mass transfer and associated physical mechanism
which lead to mass and angular momentum loss. We use a formula
based on that used to calculate angular momentum loss via a
stellar wind (Paczy\'nski \& Zi\'o\l kowski 1967; Zi\'o\l kowski
1985 and De Greve 1993). We can express the change in the total
orbital angular momentum ($J$) of a binary system as

\begin{equation}
\frac{\dot{J}}{J} = f_{1} \: f_{2} \: \frac{M_{\rm 1} \:
{\dot{M}}_{\rm 2}}{M_{\rm 2} \: M_{\rm tot}},
\end{equation}

where, $M_1 $, $M_2 $ and $M_{tot} $ denote, respectively, the
mass of the primary, secondary and the total mass of the system,
$f_{1}$ is the ratio of the mass ejected by the wind to that
accreted by the primary component and $f_2$ is defined as the
effectiveness of angular momentum loss during mass transfer (Sarna
\& De Greve 1994, 1996).

Models of secondary stars filling their Roche lobes were computed
using a standard stellar evolution code based on the Henyey-type
code of Paczy\'nski (1970), which has been adapted to low-mass
stars (as described in detail in Marks \& Sarna 1998). We use the
Eggleton (1983) formula to calculate the size of the secondary's
Roche lobe.

For radiative transport, we use the opacity tables of Iglesias \&
Rogers (1996). Where the Iglesis \& Rogers (1996) tables are
incomplete, we have filled the gaps using the opacity tables of
Huebner et al. (1977). For temperatures lower than 6000 K, we use
the opacities given by Alexander \& Ferguson (1994) and Alexander
(private communication).

To understand the evolution of close binary V228 we computed
various evolutionary sequences: for different chemical
compositions Z=0.006--0.2; initial secondary masses
0.85--1.35$M_\odot $ and initial mass ratios ($q_i =
M_{1,i}/M_{2,i}$) from 0.6 to 0.95. For each system the secondary
fills Roche lobe with a small helium core (Hertzsprung gap).

The lower conservative limit for total mass of the system is about
1.7$~M_\odot$, which infer that the original primary had a mass
exceeding 0.85$~M_\odot $.

\section{Evolutionary status -- results}

From computed evolutionary sequences we predict that:

\noindent 1. Initial mass of the primary and secondary was about
0.85--0.9$~M_\odot $, and mass ratio around 1;

\noindent 2. Current properties of the system indicate that the
original primary filled Roche lobe in the Hertzsprung gap -- early
case B mass transfer;

\noindent 3. The initial parameters for the evolutionary model are
as follow: $M_{1,i} = 0.88~M_\odot $, $M_{2,i} = 0.85~M_\odot $,
$P_i=1.35 ~$days, $f_1$=0.05, $f_2$=4.65 and Z=0.006
([Fe/H]=--0.67);

\noindent 4. The best fitting model implies loss of about 50 per
cent of initial total orbital momentum, but only 5 per cent loss of
initial total mass;

\noindent 5. The less massive component have a small helium core
of mass 0.12--0.17$~M_\odot $ and exchange mass in the nuclear
time scale;

\noindent 6. The best fitting model (0.88+0.85) spend about 10 Gyr
in detached configuration, while 0.2--0.3 Gyr in semidetached.

\section{Discussion}

V228 is eclipsing binary system of the globular cluster 47 Tuc. An
analysis of Ka\l u\.zny et al. (2007) shows that V228 is a
semidetached, low---mass Algol type binary with the less massive
component filling its Roche lobe (mass transfer occurs on the
nuclear timescale). According to accepted evolutionary scenarios,
the Algol--type binaries forms by mass transfer from initially more
massive star to less massive, leading to the reversal of the initial
mass ratio of the binary.  This provide a unique opportunity to look
into stellar interiors and to observer the remnant products of past
core hydrogen burning. The current secondary component of V228 is
oversized and overluminous for its mass, which suggest a small mass
helium core inside it. Several well--studied systems of this kind
are known in the literature. In Table 2 we summarized some orbital
and physical parameters for low--mass Algol--type binaries with
orbital period shorter than 10 days.

\begin{table*}[h!]
\begin{center}
\begin{tabular}{lllrccccl}
\multicolumn{9}{l}{Table 2 ~~The orbital and physical parameters for the Algol--type systems} \\
\hline
Name& $\rm P_{orb}$ [d] & $\rm M_1/M_\odot$ & $\rm  M_2/M_\odot$ & $\rm R_1/R_\odot$ &
$\rm R_2/R_\odot$ & $\rm f_1$ & $\rm f_2$ & References \\
\hline S Cnc & 9.48 & 2.40 & 0.20 & 2.18 & 4.83 & 0.09 & 4.62 & 1, 2, 11\\
AS Eri & 2.66 & 1.92 & 0.21 & 1.57 & 2.25 & & & 3, 4 \\
RY Aqr & 1.97 & 1.26 & 0.26 & 1.28 & 1.79 & & & 5, 6 \\
R CMa  & 1.14 & 1.07 & 0.17 & 1.50 & 1.15 & & & 7, 8, 9 \\
V228   & 1.15 & 1.51 & 0.20 & 1.36 & 1.24 & 0.05 & 4.65 & 10, see
text\\
\hline \multicolumn{9}{l}{(1) Popper \& Tomkin (1984); (2) Van Hamme
\& Wilson (1993); (3) Popper (1980); (4) Van Hamme}\\

\multicolumn{9}{l}{\& Wilson (1984); (5) Helt (1987); (6) Popper
(1989); (7) Tomkin (1985); (8) Sarma et al. (1996);}\\

\multicolumn{9}{l}{(9) Varricatt \& Ashock (1999); (10) Ka\l u\.zny
et al. (2007); (11) Sarna \& De Greve (1996).}\\

\end{tabular}
\end{center}
\end{table*}

As a result of recent analysis by Sarna et al. (1997, 1998), we
should concluded that the physical conditions in mass lossing star
of mass ranging from 0.2 to 1.5$~\rm M_\odot$ that have undergone
low--mass Algol--type evolution are most favourable for the
occurrence of intensive magnetic braking (Skumanich 19972; Verbunt
\& Zwaan, 1981). In fact during evolution of S Cnc and V 228
binaries, systems loses only less than 5 and 9\% (see table 2) of
its total mass, respectively, but about more than 50\% of its
initial total angular momentum (Sarna \& De Greve 1996). This lead
to the conclusion that a possibility of cyclic dynamos is a more
quantitative way to explain the observed status of systems presented
in Table 2.


\section*{\sc Acknowledgements}

This work were supported by grands 1 P03D 00128 and
76/E--60/SPB/MSN/P--03/DWM35/2005--2007 from the Ministry of
Sciences and Higher Education.


\end{document}